# List of main changes

1) We have explicitly written the formula of the effective area $A_{\text{DFG}}$ and we have pointed out the limit of validity of our formalism in bulk configurations.

2) We have limited the analysis of the phase mismatched system to the R=1 case(we have eliminated former Fig. 3).

3) We have added a comment on the possible useful applications of our charts in designing an experimental set-up. Their use is very practical, since require just the calculations of two numbers that take into account all the involved physical quantities.

4) We have substituted former Fig. 3 with a plot showing our best fit of Ding's scaling data. This shows that our model works much better than the linear one proposed by Ding

5) We have changed the comparison with the Ding's data, adding also explicit formulas for the calculations of the overlap integral

6) We have added a conclusion to summarize the presented results and achievements.

# Universal charts for optical difference frequency generation in the terahertz domain

Matteo Cherchi,[1,2,a)] Saverio Bivona,[1,2] Alfonso C. Cino,[3] Alessandro C. Busacca,[1,3] and Roberto L. Oliveri[1,2]

[1]*CNISM – Consorzio Nazionale Interuniversitario per le Scienze Fisiche della Materia, Unità di ricerca di Palermo, Università di Palermo, Piazza Marina 57, I-90133 Palermo*

[2]*DIFTER – Dipartimento di Fisica e Tecnologie Relative, Università di Palermo, viale delle Scienze edificio 18, I-90128 Palermo, Italy*

[3]*DIEET- Dipartimento di Ingegneria Elettrica, Elettronica e delle Telecomunicazioni, Università di Palermo, viale delle Scienze edificio 9, I-90128 Palermo, Italy*

We present a universal and rigorous approach to study difference frequency generation in the terahertz domain, keeping the number of degrees of freedom to a minimum, through the definition of a suitable figure of merit. The proposed method relies on suitably normalized charts, that enable to predict the optical-to-terahertz conversion efficiency of any system based on wave propagation in quadratic nonlinear materials. The predictions of our approach are found to be in good agreement with the best experimental results reported to date, enabling also to estimate the $d_{22}$ nonlinear coefficient of high quality GaSe.

a) cherchi@unipa.it

Difference Frequency Generation (DFG), is one of the most promising physical mechanism to generate terahertz radiation from optical sources [1], [2]. It exploits the quadratic nonlinear susceptibility of quadratic nonlinear materials to convert optical pump photons with frequency $\omega_u$ into optical signal photons with frequency $\omega_v < \omega_u$ and terahertz photons of frequency $\omega_w = \omega_u - \omega_v$. Experimental results reported to date [3] [4] lie well below the quantum efficiency limit and can be easily analyzed in terms of the small conversion approximation [5]. Only recently Ding et al. [6] experimentally achieved 39.2% photon conversion efficiency in a GaSe crystal, and we predicted more than 40% efficiency in guided wave configurations [7]. In this regime the exact solution of the coupled equations governing the process is mandatory.

DFG experiments can be performed either in free space or in waveguides, in phase mismatch or in phase matching. In all cases it is possible to treat the system with a scalar model, by defining a suitable effective nonlinear coefficient $d_{\text{eff}}$ [5] and an effective area $A_{\text{DFG}} \equiv \langle e_w | e_w \rangle \langle e_v | e_v \rangle \langle e_u | e_u \rangle / \langle e_w e_v | e_u \rangle^2$, that is the inverse of the overlap integral of the spatial distributions $e_q(x,y)$ of the three waves [8]. In this way it is possible to define a Figure of Merit [7] (FOM) $\mathcal{F} \equiv \xi^2/\alpha^2$ (having the dimensions of the inverse of a photon flux) featuring the terahertz absorption coefficient $\alpha$ and the coupling coefficient $\xi \equiv d_{\text{eff}} [2Z_0 \hbar \omega_w \omega_u \omega_v/(c^2 n_w n_u n_v A_{\text{DFG}})]^{1/2}$, where $c$ is the vacuum speed of light, $\hbar$ is the Planck constant, $Z_0$ is the vacuum impedance, $n_q$ ($q = u, v, w$) are the refractive indexes for the three waves. This enables to write the coupled equations in terms of the normalized distance $\zeta \equiv z\alpha$, of the normalized momentum mismatch $\kappa \equiv 2\Delta k/\alpha$ and of the normalized photon fluxes $\hat{q}(z;t) \equiv \sqrt{\mathcal{F}}\, q(z;t)$, as follows

$$\begin{cases} \dfrac{d\hat{w}}{d\zeta} = -i\hat{u}\hat{v}^* - \dfrac{1-i\kappa}{2}\hat{w} \\ \dfrac{d\hat{v}}{d\zeta} = -i\hat{u}\hat{w}^* \\ \dfrac{d\hat{u}}{d\zeta} = -i\hat{v}\hat{w} \end{cases} \qquad (1)$$

Here we are assuming terahertz absorption lengths much longer than terahertz wavelengths [10], negligible optical losses, and pulse durations not too smaller than the time of flight in the system, in order to avoid the effects of group velocity dispersion. The space-time dependent $q(z;t)$ functions are assumed to be slowly varying functions of $z$ and are normalized such that their square moduli are the photon fluxes $N_q$ (number of photons per unit time) of each wave. While Eqs. (1) hold exactly for guided wave experiments, where $A_{DFG}$ is constant, in the case of free space experiments, they can reliably model nonlinear conversion only if the terahertz beam profile can be assumed to be almost constant along the whole crystal length $L$, that is if the waists of the optical beams are not too smaller than $\sqrt{2cL/(n_w\omega_w)}$. Otherwise, these equations overestimate terahertz conversion, and their predictions can provide just an upper limit to maximum conversion efficiency.

A closed-form solution for equations (1) is not available but in the unrealistic cases of negligible terahertz losses or equal losses in all of the three modes [10]. Anyway, in this dimensionless form, the number of independent variables for terahertz generation (i.e. with initial terahertz photon flux $N_{w0} = 0$) is reduced to four. They are: the initial normalized pump photon flux $\hat{N}_{u0} \equiv |\hat{u}_0|^2 = \mathcal{F}N_{u0}$, the ratio $R \equiv N_{v0}/N_{u0}$ between the initial signal and pump photon fluxes, the normalized phase mismatch $\kappa$, and the normalized propagation distance $\zeta$. Notice that, since $\hat{N}_{w0} = 0$, the number of generated terahertz photons doesn't depend on the initial phases of the optical pump and of the optical signal. In general, by fixing a constraint to any two of the aforementioned four variables, it is possible to plot universal charts for the terahertz photon conversion efficiency $\eta(N_{u0},R,\kappa,\zeta) \equiv N_w/N_{u0}$ as a family of curves that are all functions of one of the two remaining variables, each curve corresponding to a different value of the other unconstrained variable, acting as a free parameter. For the sake of practice, it is also convenient to introduce a reference pump power $\bar{P}_u \equiv \hbar\omega_u/\mathcal{F}$, in order to define the normalized input power $\hat{P}_{u0} \equiv P_{u0}/\bar{P}_u = \hat{N}_{u0}$.

We now focus on phase matched DFG. In Fig. 1.a we set $R = 1$, to plot the conversion efficiency $\eta$ vs. the normalized length $\zeta$ for different $\hat{P}_{u0}$ values. The same is shown in Fig. 1.b for $R = 10^{-2}$. It is clear that, in all cases, there is one and only one propagation distance $\zeta_{max}$ corresponding to a maximum conversion efficiency $\eta_{max}$. In Fig. 1.c and Fig. 1.d we set the constraint for $\zeta$ to correspond to these maximum conversion efficiency points and we plotted the corresponding $\zeta_{max}$ and $\eta_{max}$ values as a function of $\hat{N}_{u0}$, treating $R$ as a parameter. By looking at Fig. 1.a, it s clear that, in all cases, $\eta$ initially grows with the square of $\zeta$, until the propagation length approaches $\zeta = 1$, i.e. the absorption length. Then, if $P_{u0} << \bar{P}_u$, the conversion process enters a regime where terahertz generation exactly counterbalance terahertz absorption, so that the conversion efficiency is almost constant. In this regime the pump field acts as an energy reservoir until all pump photon are converted, so that $\hat{N}_w$ and $\hat{N}_u$ are doomed to decay exponentially. Since the proposed model holds only when $1/\alpha \geq 1$ mm, in all practical cases sample lengths will not exceed hundreds of absorption lengths. Also, for very long sample, a more realistic analysis should also take into account optical losses. For higher initial powers the conversion efficiency is higher and the energy reservoir is exhausted earlier. In particular, when the depleted pump regime is reached at lengths smaller or comparable with $\alpha^{-1}$, the plateau disappears and it is replaced by an appreciable damped oscillating behaviour, due to back conversion of terahertz photons into pump photons. On the other hand, from Fig. 1.b, it is clear that the conversion dynamics is different when $R \ll 1$. In this case, for $P_{u0} \ll \overline{P}_u$, after the quadratic growth and the plateau, there is a regime where the amplified signal power becomes comparable to the pump power, so that $\eta$ starts growing faster, until pump depletion. Again, with higher pump powers the plateau regime is shorter, and the oscillating behaviour becomes appreciable. So, when $P_{u0} \ll \overline{P}_u$, for $R = 1$ maximum efficiency $\eta_{max}$ occurs at the beginning of the plateau, when the pump is almost undepleted, while for $R = 10^{-2}$ it occurs after the plateau, when pump power is halved. This is highlighted in Fig. 1.c, where it is clear that, for small $R$ values, $\eta_{max}$ occurs at much longer lengths $\zeta_{max}$. Also it is clear from Fig. 1.d that, for $R \ll 1$, $\eta_{max}$ is almost independent of the order of magnitude of $R$, even though $\zeta_{max}$ clearly depends on it.

In Fig. 2.a we show the effects of phase mismatch in the case of $R = 1$ and $\hat{P}_{u0} = 10^{-2}$. Notice that the typical oscillations of phase mismatched phenomena are damped for $\zeta \gg 1$ to reach a plateau level. For small conversion efficiencies, $\zeta_{max}$ doesn't depend on $\hat{P}_{u0}$, and the greater the phase mismatch the shorter the optimum length, as clearly shown in Fig. 2.c. For $\hat{P}_{u0} = 1$ (Fig. 2.b) the oscillations occur earlier. From Fig. 2.c and Fig. 2.d it is clear that the smaller $\hat{P}_{u0}$ the longer $\zeta_{max}$ and so the greater the detrimental effects of phase mismatch.

All the presented universal charts show how the proposed normalization allow to keep the number of degrees of freedom to a minimum and to analyze the contributions of every meaningful physical parameter in a very general way. They can be effectively used to design future experiments meant to approach the photon conversion limit, by simply calculating the reference power $\bar{P}_u$ (that is the FOM) of the system, and taking into account the terahertz absorption length. Starting from this two numbers, it will be easy to determine optimal optical peak powers, optimal sample lengths and the expected maximum conversion efficiency. Furthermore, these charts can be easily extended to include also the effects of the absorption $\alpha_O$ in the optical domain. More noticeably, they can be extended to include other nonlinear effects that can compete with DFG, especially when launching high optical intensities [11], [12]. A detailed analysis of these additional effects will be presented elsewhere [13].

We compare now the predictions of our formalism with the only experimental result approaching the quantum efficiency limit reported in the literature [6]. This was achieved with birefringent phase matching by suitably launching 300 kW peak power pump pulses with 1064 nm wavelength and 400 kW peak power optical signal pulses (power ratio $R=1.44$) in a 4.7 cm long GaSe crystal. The waist of the collimated Gaussian pump (*o* polarized) and signal (*e* polarized) beams were $r_u = 0.75$ mm and $r_v = 1.93$ mm respectively, corresponding to a terahertz waist $r_w = 1/\sqrt{1/r_u^2 + 1/r_v^2} = 0.70$ mm and to an effective area $A_{DFG} \approx 6.8$ mm$^2$. The output *e* polarized terahertz wave at 1.48 THz had a peak power of 389 W, corresponding to an external phase matching angle θ ≈ 10°. Taking into account the Fresnel reflection coefficients for all the three waves, this corresponds to a photon conversion efficiency inside the crystal of about 39.2%. High quality GaSe has an optical absorption coefficient $\alpha_O \approx 0.13$ cm$^{-1}$ at 1064 nm [14] and also a very low terahertz absorption coefficient $\alpha = 0.2$ cm$^{-1}$ at 1.48 THz [6], corresponding to a normalized sample length $\zeta = 0.94$ and to a normalized optical absorption coefficient $\hat{\alpha}_O \equiv \alpha_O / \alpha = 0.65$. The linear fit of terahertz conversion vs. sample length proposed in Ref. [6]

is both unphysical and quite unsatisfactory. Instead we present in Fig. 3 a very good fit based on our formalism, corresponding to $\hat{P}_{u0} = \hat{N}_{u0} = 1.2$, that is to a nonlinear coefficient $d_{22} \approx 43$ pm/V. As it should be expected, this value is lower than the 75 pm/V value reported for these high quality crystals [14] in the optical domain, as a result of the interplay between electronic and ionic contributions [7,15]. Small discrepancy in the fit are ascribable to different crystal quality of the samples. Also, from Fig. 3 it is clear that conversion efficiency cannot be improved further with longer GaSe samples, but only by launching higher optical peak intensities or by improving further the material properties.

To conclude, we have proposed a novel tool to design DFG experiments in the terahertz domain. We expect the proposed charts to be very helpful to determine the optimal sample length and the optimal optical peak intensities for any chosen configuration, in particular in the regime of high pump depletion. As a matter of fact, relying on this formalism, we have been able to give a satisfactory interpretation of the only experimental results approaching the quantum limit available in the literature.

M. Cherchi was supported by the Consorzio Nazionale Interuniversitario per le Scienze Fisiche della Materia (CNISM).

# Figure legends

1. Universal charts of phase matched systems. Conversion efficiency $\eta$ vs normalized propagation length $\zeta$ (a) for $R=1$ and (b) for $R=10^{-2}$. Each curve corresponds to a different normalized pump peak power $\hat{P}_{u0}$. (c) Optimum conversion length $\zeta_{\max}$ vs. $\hat{P}_{u0}$ and (d) corresponding maximum conversion efficiency $\eta_{\max}$. Each curve corresponds to a different $R$ value.

2. Universal charts of phase mismatched systems with $R=1$. Conversion efficiency vs normalized propagation length (a) for $\hat{P}_{u0}=1$ and (b) for $\hat{P}_{u0}=10^{-2}$. (c) Optimum conversion length $\zeta_{\max}$ vs. $\hat{P}_{u0}$ and (d) corresponding maximum conversion efficiency $\eta_{\max}$. Each curve corresponds to a different normalized phase mismatch $\kappa$.

3. Fit of the conversion efficiency scaling with length based on the experimental data obtained in different GaSe crystals (crosses). The only fit parameter is the nonlinear coefficient $d_{22}$, that is the normalized power value $\hat{P}_{u0}$.

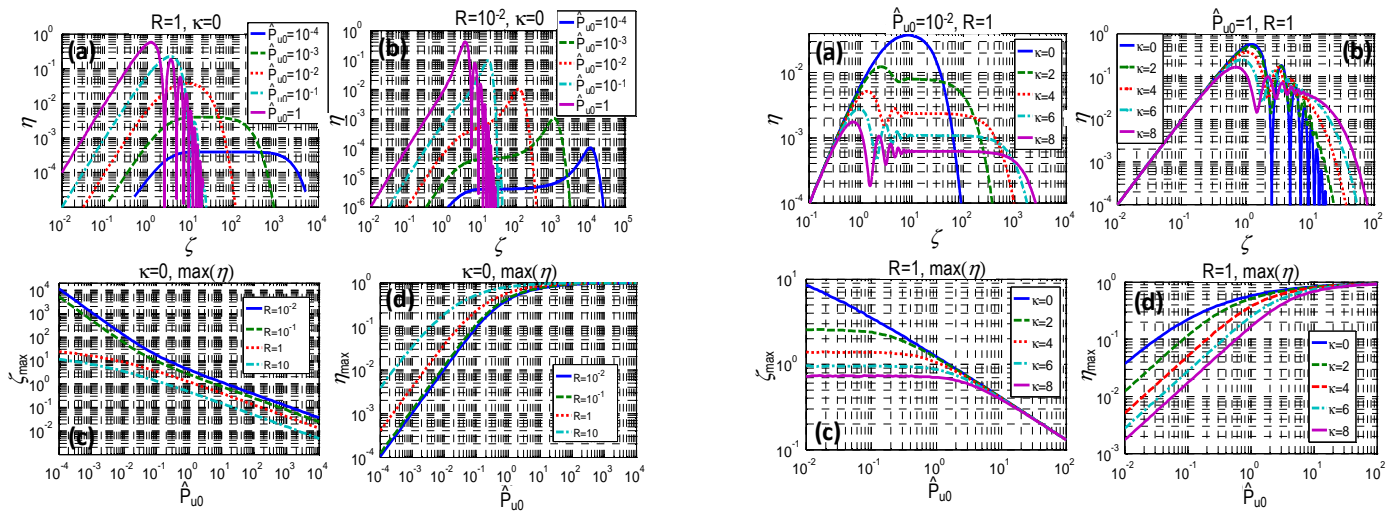

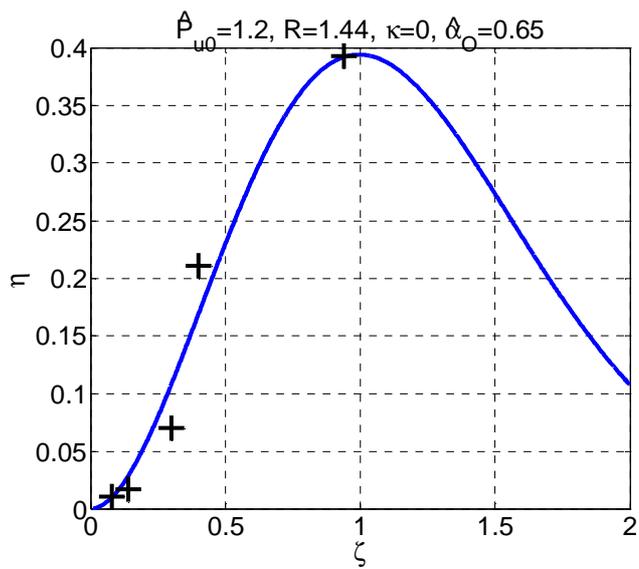